\documentclass{spie}

\usepackage{microtype}
\usepackage[separate-uncertainty]{siunitx}
\usepackage{graphicx}
\usepackage[hidelinks]{hyperref}
\usepackage{fullpage}
\usepackage{textgreek}

\usepackage[hang,perpage]{footmisc}

\renewcommand\Affilfont{\normalsize}

\makeatletter
\renewcommand\@author{\ifx\AB@affillist\AB@empty\AB@author\else
      \ifnum\value{affil}>\value{Maxaffil}\def\rlap##1{##1}%
    \AB@authlist\\[\affilsep]\ifx\@behalf\empty\else\Affilfont\@behalf\vspace{\affilsep}\\\fi\AB@affillist
    \else  \AB@authors\fi\fi}
\def\behalf#1{\gdef\@behalf{#1}}
\behalf{}     
\makeatother

\makeatletter
\newcommand\authorcount{86}
\renewcommand\AB@authnote[1]{\ifnum\value{authors}<\authorcount\relax,\fi\textsuperscript{\normalfont#1}}

\makeatother

\begin{document}

\title{Overview and status of BICEP Array's\\BA4-90/150 CMB polarimeter}

\author[a]{\href{https://orcid.org/0000-0002-4436-4215}{M.~A.~Petroff}}%
\author[b]{P.~A.~R.~Ade}%
\author[c,d]{\href{https://orcid.org/0000-0002-9957-448X}{Z.~Ahmed}}%
\author[e]{\href{https://orcid.org/0000-0001-6523-9029}{M.~Amiri}}%
\author[a]{\href{https://orcid.org/0000-0002-8971-1954}{D.~Barkats}}%
\author[f]{\href{https://orcid.org/0000-0002-3351-3078}{R.~Basu~Thakur}}%
\author[g]{\href{https://orcid.org/0000-0001-9185-6514}{C.~A.~Bischoff}}%
\author[h]{\href{https://orcid.org/0000-0003-0848-2756}{D.~Beck}}%
\author[f,i]{J.~J.~Bock}%
\author[j]{V.~Buza}%
\author[h,c]{\href{https://orcid.org/0000-0003-4541-7080}{B.~Cantrall}}%
\author[f]{\href{https://orcid.org/0000-0002-1630-7854}{J.~R.~Cheshire~IV}}%
\author[k]{J.~Connors}%
\author[l]{\href{https://orcid.org/0000-0002-2088-7345}{J.~Cornelison}}%
\author[m]{M.~Crumrine}%
\author[f]{A.~J.~Cukierman}%
\author[k]{E.~Denison}%
\author[n]{L.~Duband}%
\author[a]{\href{https://orcid.org/0000-0002-7059-8728}{M.~A.~Echter}}%
\author[o]{\href{https://orcid.org/0009-0007-6718-1730}{M.~Eiben}}%
\author[a,p]{\href{https://orcid.org/0000-0003-4117-6822}{B.~D.~Elwood}}%
\author[f]{\href{https://orcid.org/0000-0002-3790-7314}{S.~Fatigoni}}%
\author[q]{\href{https://orcid.org/0000-0001-8217-6832}{J.~P.~Filippini}}%
\author[h]{A.~Fortes}%
\author[f]{M.~Gao}%
\author[g]{C.~Giannakopoulos}%
\author[h]{N.~Goeckner-Wald}%
\author[h]{\href{https://orcid.org/0000-0001-5268-8423}{D.~C.~Goldfinger}}%
\author[r,s]{S.~Gratton}%
\author[h]{J.~A.~Grayson}%
\author[f]{\href{https://orcid.org/0009-0003-6999-0129}{A.~Greathouse}}%
\author[a]{\href{https://orcid.org/0000-0001-9292-6297}{P.~K.~Grimes}}%
\author[e]{M.~Halpern}%
\author[c,d]{S.~Henderson}%
\author[m]{\href{https://orcid.org/0000-0002-3437-5228}{T.~D.~Hoang}}%
\author[k]{J.~Hubmayr}%
\author[f]{\href{https://orcid.org/0000-0001-5812-1903}{H.~Hui}}%
\author[h]{K.~D.~Irwin}%
\author[t]{M.~Izquierdo~Poza}%
\author[f]{\href{https://orcid.org/0000-0002-3470-2954}{J.~H.~Kang}}%
\author[t]{\href{https://orcid.org/0000-0002-5215-6993}{K.~S.~Karkare}}%
\author[f]{S.~Kefeli}%
\author[a,p]{\href{https://orcid.org/0009-0003-5432-7180}{J.~M.~Kovac}}%
\author[h]{C.~Kuo}%
\author[m,u]{\href{https://orcid.org/0000-0002-4540-1495}{K.~Lasko}}%
\author[f]{\href{https://orcid.org/0000-0002-6445-2407}{K.~Lau}}%
\author[g]{M.~Lautzenhiser}%
\author[h]{\href{https://orcid.org/0000-0001-5677-5188}{T.~Liu}}%
\author[j,v]{\href{https://orcid.org/0000-0002-1414-7236}{S.~C.~Mackey}}%
\author[m]{N.~Maher}%
\author[i]{K.~G.~Megerian}%
\author[f]{L.~Minutolo}%
\author[f]{\href{https://orcid.org/0000-0002-4242-3015}{L.~Moncelsi}}%
\author[h]{Y.~Nakato}%
\author[f,i]{H.~T.~Nguyen}%
\author[f,i]{R.~O’Brient}%
\author[a]{S.~N.~Paine}%
\author[f]{A.~Patel}%
\author[a,p]{\href{https://orcid.org/0000-0002-7822-6179}{A.~R.~Polish}}%
\author[n]{T.~Prouve}%
\author[m]{\href{https://orcid.org/0000-0003-3983-6668}{C.~Pryke}}%
\author[k]{C.~D.~Reintsema}%
\author[f]{T.~Romand}%
\author[h]{M.~Salatino}%
\author[f]{A.~Schillaci}%
\author[a]{B.~Schmitt}%
\author[m,u]{\href{https://orcid.org/0000-0001-7387-0881}{B.~Singari}}%
\author[f,i]{A.~Soliman}%
\author[a]{T.~St.~Germaine}%
\author[f]{\href{https://orcid.org/0000-0003-0260-605X}{A.~Steiger}}%
\author[f]{B.~Steinbach}%
\author[b]{R.~Sudiwala}%
\author[h,c]{K.~L.~Thompson}%
\author[b]{\href{https://orcid.org/0000-0002-1851-3918}{C.~Tucker}}%
\author[i]{A.~D.~Turner}%
\author[w]{\href{https://orcid.org/0000-0002-3942-1609}{C.~Verg\`{e}s}}%
\author[j,v]{A.~G.~Vieregg}%
\author[f]{\href{https://orcid.org/0000-0002-8232-7343}{A.~Wandui}}%
\author[i]{A.~C.~Weber}%
\author[m]{\href{https://orcid.org/0000-0002-6452-4693}{J.~Willmert}}%
\author[f,c,d]{\href{https://orcid.org/0000-0001-5411-6920}{W.~L.~K.~Wu}}%
\author[h]{H.~Yang}%
\author[j,l]{\href{https://orcid.org/0000-0002-8542-232X}{C.~Yu}}%
\author[a]{\href{https://orcid.org/0000-0001-6924-9072}{L.~Zeng}}%
\author[c]{\href{https://orcid.org/0000-0001-8288-5823}{C.~Zhang}}%
\author[f]{S.~Zhang}%
\affil[a]{Center for Astrophysics, Harvard \& Smithsonian, Cambridge, MA 02138, USA}%
\affil[b]{School of Physics and Astronomy, Cardiff University, Cardiff, CF24 3AA, UK}%
\affil[c]{Kavli Institute for Particle Astrophysics and Cosmology, Stanford University, Stanford, CA 94305, USA}%
\affil[d]{SLAC National Accelerator Laboratory, Menlo Park, CA 94025, USA}%
\affil[e]{Department of Physics and Astronomy, University of British Columbia, Vancouver, BC, V6T 1Z1, Canada}%
\affil[f]{Department of Physics, California Institute of Technology, Pasadena, CA 91125, USA}%
\affil[g]{Department of Physics, University of Cincinnati, Cincinnati, OH 45221, USA}%
\affil[h]{Department of Physics, Stanford University, Stanford, CA 94305, USA}%
\affil[i]{Jet Propulsion Laboratory, California Institute of Technology, Pasadena, CA 91109, USA}%
\affil[j]{Kavli Institute for Cosmological Physics, University of Chicago, Chicago, IL 60637, USA}%
\affil[k]{National Institute of Standards and Technology, Boulder, CO 80305, USA}%
\affil[l]{High-Energy Physics Division, Argonne National Laboratory, Lemont, IL, 60439, USA}%
\affil[m]{School of Physics and Astronomy, University of Minnesota, Minneapolis, MN 55455, USA}%
\affil[n]{Service des Basses Temp\'eratures, Commissariat \`a l'\'Energie Atomique, 38054 Grenoble, France}%
\affil[o]{Faculty of Physical Sciences, University of Iceland, 102 Reykjav\'ik, Iceland}%
\affil[p]{Department of Physics, Harvard University, Cambridge, MA 02138, USA}%
\affil[q]{Department of Physics, University of Illinois at Urbana--Champaign, Urbana, IL 61801, USA}%
\affil[r]{Centre for Theoretical Cosmology, DAMTP, University of Cambridge, Cambridge CB3 0WA, UK}%
\affil[s]{Kavli Institute for Cosmology Cambridge, Cambridge CB3 0HA, UK}%
\affil[t]{Department of Physics, Boston University, Boston, MA 02215, USA}%
\affil[u]{Minnesota Institute for Astrophysics, University of Minnesota, Minneapolis, MN 55455, USA}%
\affil[v]{Department of Physics, University of Chicago, Chicago, IL 60637, USA}%
\affil[w]{Lawrence Berkeley National Laboratory, Berkeley, CA 94720, USA}

\authorinfo{Further author information: (Send correspondence to M.~A. Petroff)\\E-mail: mpetroff@cfa.harvard.edu}

\maketitle


\begin{abstract}
The inflation paradigm postulates a period of rapid expansion in the early Universe, which would generate gravitational waves. These tensor perturbations would produce a faint B-mode signature in the polarization of the cosmic microwave background (CMB), but this signal is orders of magnitude weaker than that from the CMB's other anisotropy and that from astrophysical foregrounds. Placing more-stringent upper limits on this signal or making a definitive detection thus requires exceptional control over instrument and measurement systematics, in addition to extremely-deep maps. The fourth BICEP Array receiver, BA4-90/150, aims to build and improve upon the heritage of the field-leading BICEP series of small-aperture CMB experiments with a dichroic instrument observing in 90 and \SI{150}{\giga\hertz} bands, to advance the search for the inflationary B-mode signal. The instrument will utilize transition-edge-sensor bolometers, which will be read out using a new two-level time-division-multiplexed system and be fed via feedhorn-coupled orthomode transducers and refined cold refractive optics, with the goal of both improving systematics control and sensitivity over existing receivers. With a planned deployment to the South Pole in the 2026--27 austral summer, the instrument will occupy the fourth and final remaining slot in the BICEP Array mount, completing the phaseout of \emph{Keck} Array receivers. An overview of the BA4-90/150 receiver will be presented, along with a discussion of its current status and future plans for the instrument.
\end{abstract}


\section{Introduction}

As the hot, dense early Universe expanded and cooled, photons were eventually able to free stream, forming the cosmic microwave background (CMB) sourced from the surface of last scattering, around \num{380000} years after the Big Bang. Much can be gleaned about the early Universe from the angular power spectra of the CMB, both from its temperature spectrum and from its polarization, which is typically decomposed into an even- and odd-parity basis, i.e., E- and B-mode polarization. Previous experiments have measured the temperature and E-mode polarization to high precision, cementing \textLambda{}CDM as the standard model of cosmology.\cite{Bennett2013} However, while \textLambda{}CDM parameterizes spatial curvature, it explains neither the flatness of our observed Universe nor the homogeneity of the CMB. These features are explained by the inflation paradigm, which postulates that the Universe underwent a period of rapid expansion immediately after the Big Bang, flattening the Universe and placing the entire CMB in causal contact.\cite{Guth1981} Although there are many different inflationary theories, all include this rapid expansion, which would produce primordial gravitational waves that would imprint a B-mode polarization pattern in the CMB. This extremely-faint signal is typically parameterized as the tensor-to-scalar ratio, $r$, and is orders of magnitude weaker than that associated with the CMB's other anisotropy and astrophysical foregrounds.

The BICEP series of experiments has searched for the degree-scale B-mode polarization pattern indicative of inflation for the past two decades using small-aperture refractors observing from the South Pole, providing field-leading constraints on $r$.\cite{BK18} Observations are ongoing with the fourth-generation BICEP3 instrument \cite{B3} and three fifth-generation BICEP Array receivers,\cite{BA} as well as one remaining third-generation \emph{Keck} Array receiver. The sixth-generation BA4-90/150 instrument is intended to replace the final remaining \emph{Keck} Array receiver in the fourth slot in the BICEP Array mount. This new instrument will feature refined optics in combination with upgraded feedhorn-coupled dichroic detector modules operating in bands centered near 90 and \SI{150}{\giga\hertz} in combination with two-level fully-differential time-division multiplexed (TDM) readout, with the detector and readout upgrades developed in the context of the BA+ Project. Each of the $\sim 2000$ feedhorns will be sensitive to two orthogonal polarizations in both frequency bands, for a total of $\sim 8000$ optically-sensitive detectors, similar to the detector count of the existing BA2-150 and BA3-220/270 instruments. This receiver was previously known as ``PreSAT'' and was intended to field prototype optics, detectors, and readout developed for CMB-S4, using a \SI{100}{\milli\kelvin} adiabatic demagnetization refrigerator, in order to provide early risk retirement.\cite{Petroff2024} The cancellation of CMB-S4, while unfortunate, is allowing us to instead move forward with a more-robust configuration better aligned to the BICEP program with a \SI{300}{\milli\kelvin} base temperature, while still leveraging advancements made under the CMB-S4 effort. In particular, design changes are intended to further reduce instrumental systematics and make them easier to mitigate. While systematic errors are not currently limiting BICEP constraints, control of systematic errors will become ever more important as map sensitivity improves, and these errors become progressively more difficult to mitigate. Increasing per-pixel sensitivity is also a design goal. This new instrument will soon ship to the South Pole, for a planned deployment during the coming 2026--27 austral summer.

The remainder of this manuscript is arranged as follows. The instrument's optical designs will be discussed in Section~\ref{sec:optics}, followed by its detectors, readout, and module designs in Section~\ref{sec:drm}. We will then describe its current status and future plans in Section~\ref{sec:status}, before concluding in Section~\ref{sec:conclusions}.

\section{Optics}
\label{sec:optics}

\begin{figure}
\centering
\includegraphics[width=0.8\textwidth]{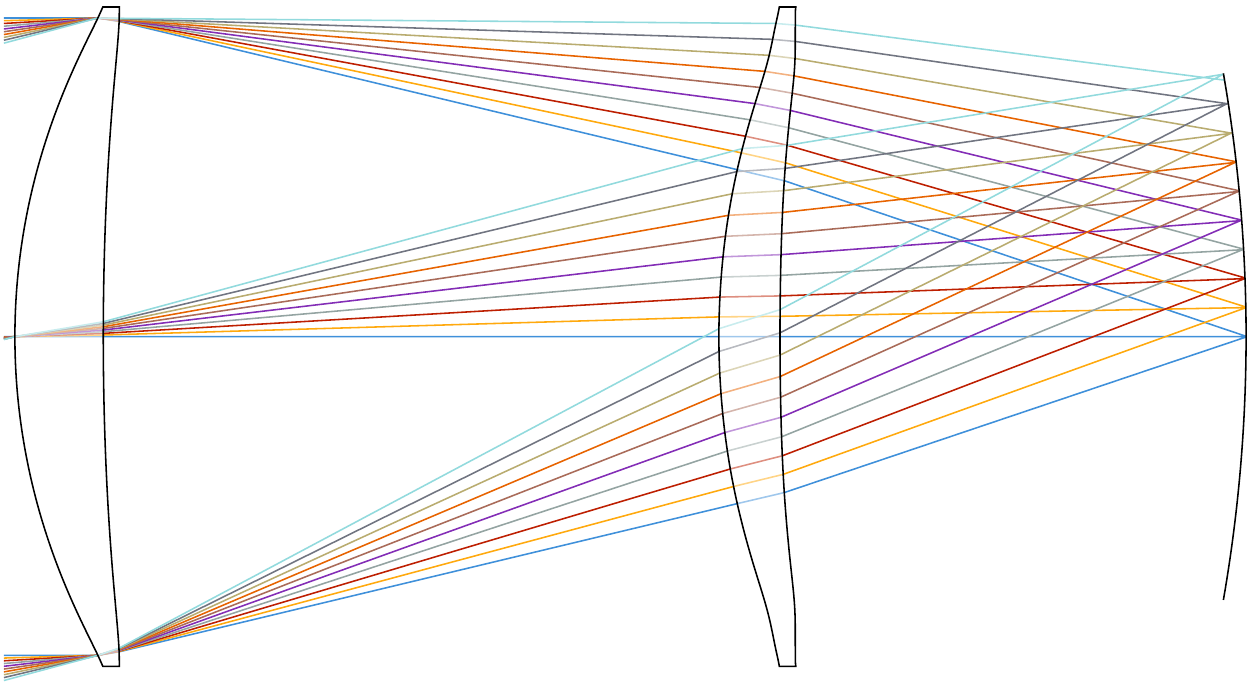}
\caption{Ray trace of the BA4-90/150 optics design, incorporating, from left to right, two high-density polyethylene (HDPE) lenses and a spherical focal surface with a radius of \SI{1400}{\milli\meter}. Flat filter elements are not included in this diagram, but their inclusion only results in slight changes to the lens placement.}
\label{fig:optics}
\end{figure}

The instrument optics use a refined prescription based on mature CMB-S4 small-aperture telescope (SAT) work, which was based on the previous BICEP Array design, which in turn was influenced by earlier CMB-S4 work. The design consists of two high-density polyethylene (HDPE) lenses and a spherical focal surface with a \SI{1.4}{\meter} radius of curvature; each lens has two aspheric surfaces, and each surface is defined by eleven even polynomial terms. Initial design work was done using Q\textsubscript{bfs} polynomials, which are an orthogonal basis,\cite{Forbes2008} as they seemed to perform better with Zemax's optimizer\footnote{Ansys, Inc., Canonsburg, PA, USA; \url{https://ansys.com/}} than more traditional even polynomials; a change of basis was then performed to even polynomials prior to final optimization, as even polynomials are significantly easier to use for mechanical drawings for manufacturing. A ray-trace of this design is shown in Figure~\ref{fig:optics}. The optics were designed using a \SI{580}{\milli\meter} aperture and \SI{30}{\degree} field of view, for a large etentue; the field of view matches prior BICEP Array receivers, but the aperture is enlarged from \SI{550}{\milli\meter}, which is possible due to other changes in the receiver to increase beam clearances. In addition to the standard terms to minimize wavefront error, the figure of merit for the optics design optimization included terms to enforce telecentricity and minimize f-ratio variation across the focal surface; these additional terms result in a design that more uniformly illuminates the aperture stop with the detector beams, in a time-reverse sense, which should reduce beam systematics. Other terms are also included to favor thinner lenses, to reduce ripple toward the edge of the lenses and enforce a monotonic surface profile, and to favor designs with rays closer to perpendicular to the lens surfaces. The final design achieves a Strehl ratio $\geq 0.99$ at \SI{150}{\giga\hertz}, with sagittal f-ratios of f/1.57--1.59 and tangential of f/1.57--1.65.

Beyond the lenses, the optics include a vacuum window and a radio-transparent multi-layer-insulation (RT-MLI) filter stack, an alumina filter, a nylon filter, and a low-pass edge filter for thermal filtering. The vacuum window is a polyethylene-composite ultra-thin window, which is $\sim$\SI{1.3}{\milli\meter} thick, similar to those previously deployed on BICEP Array receivers and BICEP3,\cite{Eiben2026} but incorporating recent process improvements that reduce the RMS thickness variation to $\sim$\SI{20}{\micro\meter} ($\sim\lambda/100$ at \SI{150}{\giga\hertz}) and with a larger clear aperture diameter of \SI{76}{\centi\meter}. The receiver is currently using an RT-MLI filter stack\cite{Choi2013} constructed from twelve \SI{3}{\milli\meter} thick layers of Zotefoams Plastazote HD30,\footnote{Zotefoams PLC, London, England; \url{https://www.zotefoams.com/}} an HDPE foam with a $\sim$\SI{400}{\micro\meter} cell size, but we are planning to switch to Zotefoams Ecozote PP35, a polypropylene foam with a $\sim$50--\SI{150}{\micro\meter} cell size,\footnote{A cell size of $\sim$\SI{50}{\micro\meter} was measured on a material sample, but Zotefoams PIS-E003 (Issue 0, Revision 4, which is provisional) specifies the cell size as \SI{150}{\micro\meter}.} for reduced in-band scattering. The \SI{10}{\milli\meter} thick alumina filter consisting of Kyocera\footnote{Kyocera Corporation, Kyoto, Japan; \url{https://global.kyocera.com/prdct/fc/}} AO479U alumina and is similar to the filters in existing BICEP Array receivers but with an enlarged diameter to increase beam clearance. The nylon 6 filter is similarly enlarged compared to prior BICEP Array receivers; it may have an identical \SI{7.5}{\milli\meter} thickness, but other thicknesses are also being evaluated. The final filtering element is a low-pass edge filter;\cite{Ade2006} we have demonstrated that we can reach base temperature on the receiver without this filter and that it is thus not necessary for cryogenic reasons, so it may be eliminated if above-band ``blue leak'' in our detectors can be sufficiently reduced without it.

In order to minimize optical systematics, all optical surfaces, with the exception of extremely-low-index foam, are anti-reflection (AR) coated, with a goal of $<0.1$\% per-surface reflection on all optical elements. For polyethylene elements, this coating will be a two-layer coating comprising an inner layer of sintered polytetrafluoroethylene (sPTFE), with an outer layer of expanded PTFE (ePTFE). The nylon filter will use a similar coating, except the sPTFE layer will be replaced with a low-density polyethylene (LDPE) layer, due to nylon's higher refractive index. For the alumina filter, a three-layer epoxy--ceramic composite coating is in development, but a fallback for the initial season may be a heritage two-layer epoxy coating comprising an inner layer of Stycast 2850FT\footnote{Henkel AG \& Co. KGaA, D\"usseldorf, Germany; \url{https://www.henkel.com/}} and an outer layer of Stycast 1090, similar to that used in the BA1-30/40 and BA3-220/270 instruments.

\section{Detectors, readout, and modules}
\label{sec:drm}

The detector modules being developed under the BA+ initiative for the receiver consist of feedhorn-coupled orthomode transducers (OMTs) feeding transition-edge sensor (TES) bolometers. Each horn includes four optically-sensitive TES bolometers, with two per orthogonal linear polarization, with pairs for both the 90 and \SI{150}{\giga\hertz} bands. Each TES includes two superconducting transitions, with a titanium ``science'' transition, as well as a higher-temperature aluminum ``lab'' / calibration transition. Between the OMTs and the TES bolometers are diplexers and on-chip band-defining filters. An additional two ``dark'' TES bolometers will be included in each pixel, although only a subset of these will be read out. The detector wafers are to be fabricated by JPL, based on prior designs for BA and CMB-S4.

Similar to prior BICEP experiments, these detectors will be read out using time-division multiplexing (TDM). This readout chain will consist of NIST mux21 SQUID multiplexing chips\cite{Durkin2023} in the detector modules, with higher gain, increased bandwidth, and reduced crosstalk compared to prior designs; NIST SA23 SQUID series arrays at \SI{4}{\kelvin}; and new warm electronics being developed at SLAC, which are being validated against the Multi-Channel Electronics (MCE) hardware developed at the University of British Columbia\cite{Battistelli2008} and are expected to have higher bandwidth. These upgraded components will use two-level row addressing for increased readout density and will enable fully-differential readout, which should reduce susceptibility to out-of-band radio-frequency interference (RFI).

The detector wafers will be integrated into a flat-pack module design, which is a hybrid of the existing BICEP Array module design and the module design developed for CMB-S4. The sky side of the module starts with a feedhorn plate, which is coupled to the detector wafer using a photonic crystal and waveguide interface structure. The detector wafer is followed by a backshort and a wiring printed circuit board (PCB) surrounded by both superconducting niobium and high-$\mu$ magnetic shielding. The SQUID multiplexing chips are installed on a set of three silicon wiring chips, which are then attached to the aforementioned PCB. The feedhorn plate includes 167 smooth-wall spline-profile feedhorns\cite{Simon2018} with an \SI{8.94}{\milli\meter} horn pitch, which will be drilled into a metal block. While aluminum is the preferred material for this block for reasons of mass and ease of fabrication, it is sub-optimal for magnetic reasons, as it concentrates the magnetic field near the TES bolometers when it becomes superconducting; thus, non-superconducting alternatives such as brass or magnesium alloy are also being considered. These modules will then be installed on copper spacers that are attached to a copper baseplate and will position the modules such that the phase centers of the feedhorns follow the spherical focal surface. In order to maximize detector packing on the focal plane, pairs of modules will be installed on these spacers, which will then be fastened to the baseplate such that the modules are only \SI{1}{\milli\meter} apart; this mounting mechanism allows for the modules to be installed without access to the bottom of the baseplate but also without requires space to fit a screwdriver or hex key in between the modules. Between four optically-sensitive bolometers per horn, 167 horns per module, and a total of twelve modules, there will be a total of $\sim$8000 optically-sensitive detectors in the receiver.

\section{Current Status and Future Plans}
\label{sec:status}

As of the time of writing---mid-to-late 2026 July---the receiver is currently cold, on its 18\textsuperscript{th} cold run. It is thus well characterized cryogenically, having productively served as a North American detector and systems testbed for the development of previous BICEP Array receivers. We plan to build up a new cryostat, designated BA5, as a testbed to serve this purpose going forward. Optical testing with \SI{90}{\giga\hertz} BICEP3 and \SI{150}{\giga\hertz} BICEP Array modules is in progress; this testing is using heritage MCE hardware for detector readout, pre-science-grade optics, and a heritage focal plane structure. Optics are in the lab, with AR coating planned in the coming weeks, as final coating materials arrive. The cold readout is half populated, with the remaining parts on order, as are the mechanical parts for the new focal plane design.

One or two more cold runs are planned for the next two months, for integration testing of optics and the focal plane. The receiver will then be packed and shipped in late September or early October, for Pole deployment this coming austral summer. The initial engineering season will include \SI{90}{\giga\hertz} (BICEP3), and possibly \SI{150}{\giga\hertz} (BA), slot-antenna-coupled modules and a NIST 90 / \SI{150}{\giga\hertz} horn-coupled module, with the latter incorporating the new mux21 SQUID chips. This will validate the new dichroic optics and allow for cross comparison of module types. We aim to populate the focal plane with BA+ modules in subsequent seasons and plan to switch to SLAC TDM warm readout electronics.

\section{Conclusions}
\label{sec:conclusions}

The BA4-90/150 receiver is intended to be a refinement and upgrade of previous BICEP Array receivers, as the BICEP collaboration's sixth-generation receiver design. Its changes are intended both to further reduce instrument systematics and to increase per-tube sensitivity. The instrument will deploy in the coming 2026--27 austral summer, replacing the last remaining \emph{Keck} Array receiver. The initial season will collect engineering data with a menagerie of detector modules for validation of the optics and comparative testing, and we intend to complete the focal plane over subsequent seasons.

\acknowledgments

We acknowledge the National Science Foundation Division of Astronomical Sciences for their support of PreSAT under Grant Number 2216223.

\bibliography{paper.bib}
\bibliographystyle{spiebib}

\end{document}